\documentclass[a4paper,11pt]{article}
\usepackage{pos}

\usepackage[normalem]{ulem}

\title{Constraints on the dark sector from electroweak precision observables}

\author*[a]{B.~M.~Loizos}
\author[a]{X.~G.~Wang}
\author[a]{A.~W.~Thomas}
\author[a]{M.~J.~White}
\author[a]{A.~G.~Williams}

\affiliation[a]{CSSM and ARC Centre of Excellence for Dark Matter Particle Physics,
Department of Physics, The University of Adelaide, Adelaide, Australia}

\emailAdd{bill.loizos@adelaide.edu.au}
\emailAdd{xuan-gong.wang@adelaide.edu.au}
\emailAdd{anthony.thomas@adelaide.edu.au}
\emailAdd{martin.white@adelaide.edu.au}
\emailAdd{anthony.williams@adelaide.edu.au}

\abstract{
We revisit the constraints on the dark photon mixing parameter from electroweak precision observables by using the latest experimental data.
We then investigate the impact of the $W$ boson mass from the CDF measurement. With connection to the dark matter sector, we extend the previous work by placing constraints directly on the coupling of the dark photon to fermionic dark matter particles. 
}

\FullConference{The XVIth Quark Confinement and the Hadron Spectrum Conference (QCHSC24)\\
 19-24 August, 2024\\
 Cairns Convention Centre, Cairns, Queensland, Australia\\}

\begin{document}
\maketitle

\section{Introduction}
The dark photon is a popular candidate when considering models that bridge ordinary matter to the dark sector. It is introduced as an extra $U(1)$ gauge boson~\cite{Fayet:1980ad, Fayet:1980rr, Holdom:1985ag, Okun:1982xi} that kinetically mixes with the Standard Model (SM) hypercharge before electroweak symmetry breaking. With the inclusion of a dark photon, the $Z$ boson mass and its weak couplings will be shifted with respect to their SM predications.  As a result, a bunch of electroweak precision observables (EWPO) will be modified by the dark photon~\cite{Hook:2010tw, Curtin:2014cca,Harigaya:2023uhg,BENTO2024138501,Davoudiasl:2023cnc}.

 The dark photon may also couple to dark matter particles $\chi$ with mass $m_{\chi}$ and interaction strength $g_{\chi}$. 
 Popular scenarios include Dirac, pseudo-Dirac, scalar, asymmetric dark matter~\cite{Izaguirre:2015yja}.
 The dark matter relic density and the dark matter direct detection could place lower and upper bounds on the variable $y = \epsilon^2 \alpha_D (m_{\chi}/m_{A'})^4$, respectively, where $\alpha_D = g^2_{\chi}/4\pi$. 
 It would be also important to set upper limits directly on $g_{\chi}$. 
 
 This contribution is based on our recent work~\cite{Loizos:2023xbj}. We first revisit the previous EWPO constraints on the dark photon parameters by using the latest experimental data, especially the new $W$ boson mass from the CDF measurement~\cite{CDF:2022hxs}. Then, we consider the case of the dark photon coupling to a dark Dirac fermion $\chi$, and place direct constraints on the coupling $g_{\chi}$.

\section{The Dark Photon Formalism}
\label{sec:sec2}

The dark photon is introduced as an additional $U(1)$ gauge boson\cite{Fayet:1980ad, Fayet:1980rr, Holdom:1985ag}, which interacts with SM particles through kinetic mixing with hypercharge~\cite{Okun:1982xi},

\begin{equation}
\label{eq:L}
 {\cal L}  \supset  
- \frac{1}{4} F'_{\mu\nu} F'^{\mu\nu} + \frac{1}{2} m^2_{A'} A'_{\mu} A'^{\mu} 
+ \frac{\epsilon}{2 \cos\theta_W} F'_{\mu\nu} B^{\mu\nu} + g_{\chi} \bar{\chi} \gamma^{\mu} \chi A'_{\mu} \, .
\end{equation}

where $F'_{\mu\nu}$ denotes the dark photon field strength tensor,  $B^{\mu\nu}$ is the SM hypercharge field strength tensor, and $\theta_W$ is the weak mixing angle. Here, $\epsilon$ represents the kinetic mixing parameter between the dark photon and the SM hypercharge gauge field. The fields $A'$ and $\bar{Z}$ correspond to the unmixed dark photon and SM neutral weak gauge boson, respectively. Note that we also introduced minimal coupling of $A'$ to dark fermions $\chi$.

By performing field redefinitions and diagonalising the mass-squared matrix, the physical $Z$ and dark photon $A_D$ are
\begin{eqnarray}
\label{eq:Z-AD}
Z_{\mu} &=& \cos\alpha \bar{Z}_{\mu} + \sin\alpha A'_{\mu} \, , \nonumber\\
A_{D\mu} &=& - \sin\alpha \bar{Z}_{\mu} + \cos\alpha A'_{\mu}\, .
\end{eqnarray}
The $\bar{Z}-A'$ mixing angle $\alpha$ is given by~\cite{Kribs:2020vyk}
\begin{eqnarray}
\tan \alpha &=& \frac{1}{2\epsilon_W} \Big[ 1 - \epsilon^2_W - \rho^2 - {\rm sign}(1-\rho^2) \sqrt{4\epsilon_W^2 + (1 - \epsilon_W^2 - \rho^2)^2} \Big] \, ,
\end{eqnarray}
where
\begin{eqnarray}
\label{eq:ew-rho}
\epsilon_W &=& \frac{\epsilon \tan \theta_W}{\sqrt{1 - \epsilon^2/\cos^2\theta_W}} ,\nonumber\\
\rho &=& \frac{m_{A'}/m_{\bar{Z}}}{\sqrt{1 - \epsilon^2/\cos^2\theta_W}} \, .
\end{eqnarray}

The tree level couplings of the physical $Z$ boson to SM fermions can be found in Refs.~\cite{Loizos:2023xbj, Kribs:2020vyk}, and the radiative corrections were given in Ref.~\cite{Cho:1999km}.
In addition, the $Z$ will also couple to the dark matter particle $\chi$, with the physical coupling given by
\begin{equation}
C_{Z,\chi\bar{\chi}} = \frac{g_{\chi} \sin\alpha}{\sqrt{1- \epsilon^2/\cos^2\theta_W}}\, .
\end{equation}

\section{Electroweak precision observables}

The electroweak precision observables are listed in the first column of Tab.~\ref{tab:SM-fit}.  It is convenient to choose the following set of free parameters
\begin{equation}
\label{eq:parameters}
    m_h, m_{\bar Z}, m_t, \alpha_s, \Delta\alpha^{(5)}_{\rm had}\, .
\end{equation}
The dependences of the $Z$ pole observables on these parameters and the dark photon modifications are encoded in the physical couplings of the $Z$ boson~\cite{Curtin:2014cca, Loizos:2023xbj}.
For the $W$ boson mass and its decay width, we use the parametrisations in Ref.~\cite{Awramik:2003rn} and Ref.~\cite{Cho:2011rk}, respectively.

Following the procedure performed in Ref.~\cite{Curtin:2014cca}, we define
\begin{equation}
    \chi^2= V \cdot cov^{-1} \cdot V\, ,\ \ cov = \Sigma_{\rm exp} \cdot cor \cdot \Sigma_{\rm exp}\, ,
\end{equation}
where $V = {\rm theory} (m_h,m_Z,m_t,\alpha_s,\Delta\alpha^{(5)}_{\rm had}, m_{A_D}, \epsilon,  g_{\chi}) - {\rm exp}$ denotes the difference vector between the theoretical predictions and the experimental measurements. The experimental data used are from the latest dataset found in~\cite{ParticleDataGroup:2022pth}, detailed in the second column of Tab.~\ref{tab:SM-fit}, with the $Z$ pole observables taken from Ref.~\cite{Janot:2019oyi}, using an improved Bhabha scattering cross section. Additionally, $m_W^{\rm PDG}$ denotes world average of the $W$ boson mass from measurements conducted at LEP, SLC, Tevatron and LHC.
The covariance matrix $\Sigma_{\rm exp}$ denotes the experimental uncertainties and is diagonal. 
The correlation matrices denote the correlations among the electroweak precision observables, and are adopted from ~\cite{Janot:2019oyi} and ~\cite{ALEPH:2005ab} for the $Z$ and quark observables, respectively. Furthermore, a correlation coefficient of $-0.174$ exists between the mass and width of the $W$ boson. 

\section{Constraints on the dark parameters}
\label{sec:sec4}

\subsection{Standard Model Fit}
We first perform a Standard Model fit to electroweak precision observables without new physics. 
The parameters in Eq.~(\ref{eq:parameters}) are varied around their measured values in order to minimise the value of $\chi^2$.
In the case of $m_W^{\rm PDG}$, the best fit results are shown in the third column of Tab.~\ref{tab:SM-fit} with $\chi^2_{\rm SM} = 12.9$. Substituting the world averaged $m_W$ with the latest CDF measurement~\cite{CDF:2022hxs} significantly increases the minimum $\chi^2$ to 68.2.

\begin{table*}[!htpb]
\begin{tabular}{cccc} \hline \hline
            Observable                   &                 Measurement             &          Fit result ($m^{\rm PDG}_W$)      &    Fit result ($m^{\rm CDF}_W$)     \\
  \hline
         $ m_Z$ (GeV)                  &         $91.1875\pm 0.0021$       &   91.1880 $\pm$ 0.0020      &   91.1910 $\pm$ 0.0020    \\ 
         $m_h$ (GeV)                   &                125.25 $\pm$ 0.17      &    125.25 $\pm$ 0.17           &    125.24 $\pm$ 0.17     \\
         $m_t$ (GeV)                    &              172.69 $\pm$ 0.30       &   172.75 $\pm$ 0.30           &   173.09 $\pm$ 0.29     \\
 $\alpha_s(m_Z^2)$                   &  0.1179 $\pm$ 0.0009  &    0.1203 $\pm$ 0.0026     &    0.1176 $\pm$ 0.0026     \\
$\Delta\alpha^{(5)}_{\rm had}$  &         $0.02757 \pm 0.00010$         &     0.02755 $\pm$ 0.00010      &     0.02745 $\pm$ 0.00010    \\ \hline
    $\Gamma_Z$ (GeV)             &         2.4955 $\pm$ 0.0023          &          2.4963                       &     2.4953          \\
$\sigma_{\rm had}^0$ (nb)       &            41.4802 $\pm$ 0.0325           &         41.4704                    &      41.4814       \\
         $A_\ell$                           &                $0.1499\pm 0.0018$        &         0.1471                     &       0.1476             \\
         $A_b$                             &              $0.923\pm 0.020$              &        0.935                        &      0.935                \\
          $A_c$                             &                  $0.670\pm 0.027$         &         0.668                        &     0.668                \\
         $R_\ell^0$                      &                 20.7666 $\pm$ 0.0247         &        20.7529                      &      20.7358             \\
           $R_b^0$                       &             $0.21629 \pm 0.00066 $     &         0.21581                    &      0.21581          \\
           $R_c^0$                      &                 $0.1721 \pm 0.0030$       &        0.1723                        &       0.1722            \\
   $A_{FB}^{\ell,0}$                 &                 $0.0171\pm 0.0010$        &         0.0162                      &       0.0164          \\
   $A_{FB}^{b,0}$                   &                 $0.0992\pm 0.0016$         &        0.1031                     &       0.1034        \\              
     $A_{FB}^{c,0}$                 &                  $0.0707 \pm 0.0035$       &        0.0737                      &      0.0739          \\
  $m^{\rm PDG}_W$ (GeV)            &                80.377 $\pm$ 0.012      &        80.359                         &            \                    \\
    $m^{\rm CDF}_W$ (GeV)         &                $80.4335 \pm 0.0094$     &               \                             &   80.3686   \\
  $\Gamma_W$ (GeV)           &                $2.085 \pm 0.042$           &        2.091                          &    2.091      \\ \hline
  $\chi^2_{d.o.f}$                    &                                    \                     &     $12.92/(17-5)$    &    $68.23/(17-5)$     \\ \hline \hline
\end{tabular}
\caption{SM fit results.
The experimental value of $\alpha_s(m^2_Z)$ is not included in the fit.}
\label{tab:SM-fit}
\end{table*}

\subsection{Constraints on the dark photon}
We place constraints on the kinetic mixing parameter of the dark photon, $\epsilon$, initially neglecting its coupling to dark fermions by setting $g_\chi = 0$. For chosen dark photon mass $m_{A_D}$, adjust $\epsilon$ and repeatedly fit to the electroweak precision observables, whilst allowing the parameters in Eq.~(\ref{eq:parameters}) to float freely. The resulting minimum $\chi^2$ is dependent upon $\epsilon$, and the 95\% confidence level exclusion region is defined such that: 
\begin{equation}
\label{eq:chi2_AD}
    \chi^2_{A_D} (\epsilon) - \chi^2_{\rm SM} \ge 3.8 \, .
\end{equation}

\begin{figure*}[!t]
\begin{center}
\includegraphics[width=0.95\textwidth]{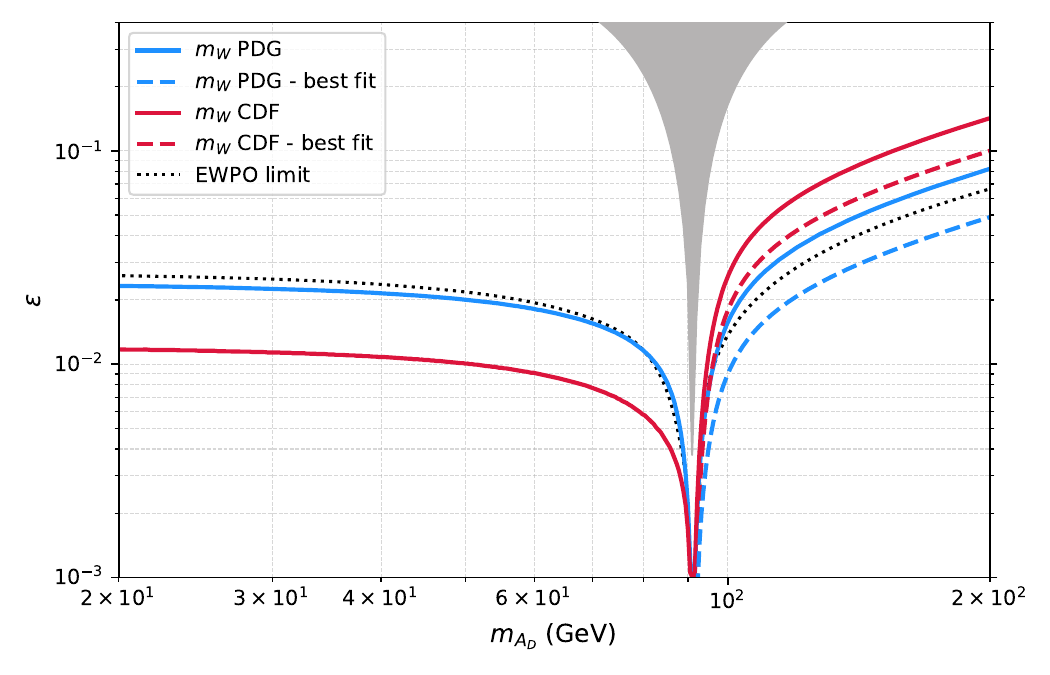}
\vspace*{-0.1cm}
\caption{ The \textit{solid} blue (red) curves represent the 95\% CL exclusion constraints on $\epsilon$ for the case in which $m_W$ is taken to be the PDG (CDF) result. The blue (red) \textit{dashed} curves represents the dark photon parameters that provide the \textit{best fit}, i.e the minimum $\chi^2_{A_D}$value that can be obtained by floating $\epsilon$, for the PDG (CDF) value of $m_W$.
The region in grey is not accessible due to the ``eigenmass repulsion" associated with the Z mass. The EWPO limit (black dotted) is taken from Ref.~\cite{Curtin:2014cca}.
The best fit curve (red dashed) is consistent with the result in Ref.~\cite{Thomas:2022gib}.}
\label{fig:EWPO-revised}
\end{center}
\end{figure*}

We present the upper limits on the kinetic mixing parameter $\epsilon$ in Fig.~\ref{fig:EWPO-revised}. Using the PDG value for $m_W$, our findings (solid blue curve) align qualitatively with previous literature~\cite{Curtin:2014cca}. We have extended this analysis by incorporating the CDF measurement of $m_W$. The analysis reveals that the constraints on $\epsilon$ become tighter for $m_{A_D} < m_Z$, and relax for $m_{A_D} > m_Z$, depicted by the solid red curve in Fig.~\ref{fig:EWPO-revised}.

Notably, when $m_{A_D} < m_Z$, including the dark photon consistently worsens the $\chi^2$ with respect to the SM fit. In Fig.~\ref{fig:EWPO-revised}, the blue dashed and red dashed lines represent the dark photon parameters corresponding the maximum $\chi^2$ reduction achieved upon fitting the PDG value and the CDF value of $m_W$, respectively. The latter result is consistent with Refs.~\cite{Thomas:2022gib, Zhang:2022nnh, Zeng:2022lkk, Cheng:2022aau}. 

For the PDG $m_{W}$ value, the fit is slightly improved above the $Z$-pole, yielding $\chi^2_{A_D}= 11.38$ at $(m_{A_D}, \epsilon) = (200\ {\rm GeV}, 0.0489)$. In the case for the CDF $m_{W}$ measurement, the fit notably improves with $\chi^2_{A_D} = 33.7$ at $(m_{A_D}, \epsilon) = (200\ {\rm GeV}, 0.1001)$. The predicted value for $m_W$ is $80.4060\ {\rm GeV}$, reducing the discrepancy to $2.9\ \sigma$.
Despite offering an improved fit compared to the Standard Model, the resulting  $\chi^2$ is still substantially large.

\subsection{Constraints on dark fermions}
When $g_\chi$ is turned on, the $Z$ boson decay width will receive additional contribution from $\chi\bar{\chi}$ final states if $m_{\chi} < m_Z / 2$,
\begin{eqnarray}
\Gamma_Z = \Gamma_{\rm had} + \Gamma_e + \Gamma_\mu + \Gamma_\tau + 3 \Gamma_{\nu} + \Gamma_{\chi}
\end{eqnarray}
where $\Gamma_{\chi}$ is the $Z\rightarrow \chi {\bar \chi}$ partial width, which depends on the dark photon mixing parameter $\epsilon$ and the coupling to dark fermions $g_{\chi}$,
\begin{eqnarray}
\label{eq:Gamma_chi}
\Gamma_{\chi} = \frac{m_Z C^2_{Z,\chi\bar{\chi}}}{12\pi} \left( 1 + \frac{2 m^2_{\chi}}{m^2_Z} \right) \sqrt{1 - \frac{4 m^2_{\chi}}{m^2_Z}}\, .
\end{eqnarray}
 This allows us to place constraints on $\epsilon$ and $g_{\chi}$ independently via electroweak precision observables, using the 95\% exclusion zone for two parameters, given by
\begin{eqnarray}
\chi^2_{A_D}(\epsilon, g_{\chi}) - \chi^2_{\rm SM} \ge 5.99\, .
\end{eqnarray}

\begin{figure*}[!htpb]
\centering
\includegraphics[width=1.06\textwidth]{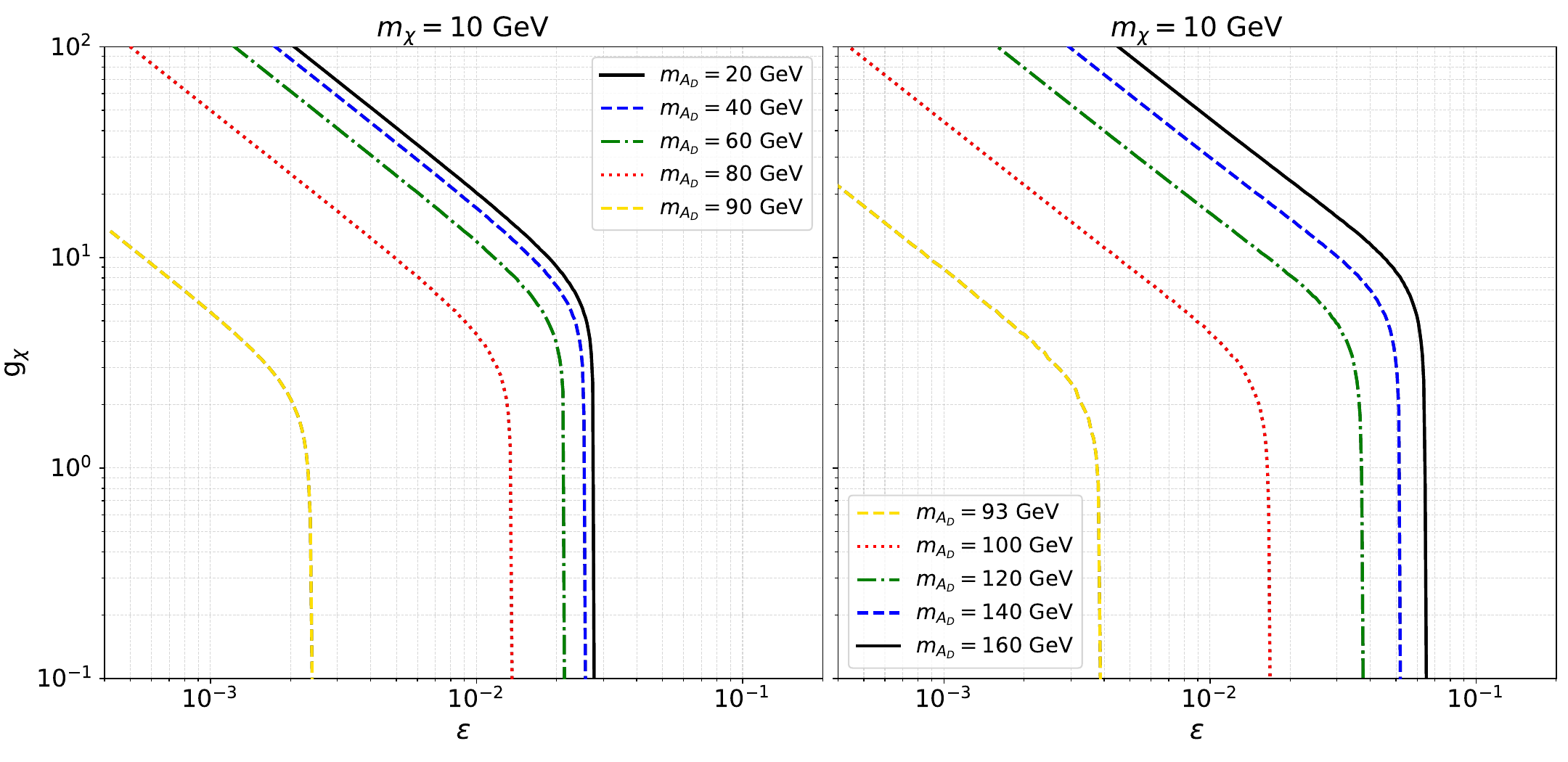}
\vspace*{-0.1cm}
\caption{ The 95\% CL exclusion constraints on dark parameters in the $g_{\chi} - \epsilon$ plane, using $m_W^{\rm PDG}$. Each curve corresponds to $\chi^2_{A_D}(\epsilon, g_{\chi}) - \chi^2_{\rm SM} = 5.99$ for each respective $m_{A_{D}}$.}
\label{fig:g_chi_eps}
\end{figure*}

In Fig.~\ref{fig:g_chi_eps}, we show results utilising the PDG value for the $W$ boson mass. We present fits for heavy dark fermions with $m_{\chi}= 10\ {\rm GeV}$ for several dark photon masses $m_{A_D}$. For dark photon masses below $m_Z$, constraints on both the mixing parameter $\epsilon$ and the dark fermion coupling $g_\chi$ strengthen as $m_{A_D}$ increases. Conversely, the constraints become relaxed for masses above $m_{Z}$.

Note that comparisons between the results in Fig.~\ref{fig:g_chi_eps} and existing constraints is non-trivial. Current astrophysical and direct detection experiments place constraints on the combination $y = \epsilon^2 \alpha_D (m_{\chi}/m_{A'})^4$ for light dark matter scenarios, which are insensitive to individual variations of $\epsilon$ and $\alpha_D$~\cite{Izaguirre:2015yja}. Typically, constraints placed on electroweak precision observables from astrophysical and collider experiments involve a scaling by $\alpha_D$, which can range from the $\alpha$ value determined through QED, up to perturbativity bound. Our analysis directly presents upper bounds on $g_\chi$, which are shown to be significantly suppressed as $\epsilon$ approaches its exclusion bound. Future work can be done to investigate the constraints on $y$, given these $\epsilon$ dependent limits on $g_\chi$.

\section{Conclusions}
\label{sec:conc}
In this work, we revisited the constraints on the dark photon's kinetic mixing parameter $\epsilon$ using electroweak precision observables (EWPO), in light of the new CDF measurement of the $W$ boson mass. We discover constraints on $\epsilon$ tighten for dark photon masses $m_{A_D} < m_Z$, and relax for $m_{A_D} > m_Z$. We also explored parameter spaces for the dark photon model aiming to enhance the agreement between theory predictions and experiment.

We also considered scenarious in whichthe dark photon couples to a new dark Dirac fermion. Due to kinetic mixing, the Z boson will inherit a new invisible decay to $\chi\bar{\chi}$, in addition to its original final states. This allows us to impose constraints on $g_\chi$ by fitting to EWPO data. These constraints are rather stringent as $m_{A_{D}}$ approaches $m_{Z}$, and become weaker as the dark fermion mass increases.

Future precision experiments from proposed collider experiments such as CEPC~\cite{CEPCStudyGroup:2018ghi}, FCC-ee~\cite{FCC:2018evy}, ILC~\cite{ILC:2019gyn} and CLIC~\cite{CLICdp:2018cto} promise to confine these constraints, and our analysis could be extended to peudo-Dirac, scalar, and asymmetric dark matter scenarios.

\section*{Acknowledgments}
This work was supported by the University of Adelaide and the Australian Research Council through the Centre of Excellence for Dark Matter Particle Physics (CE200100008). This work was also supported by an Australian Government Research Training Program Scholarship.

\bibliographystyle{JHEP}
\bibliography{bibliography.bib}


\end{document}